\documentclass{article}

\usepackage{arxiv}

\usepackage[utf8]{inputenc} % allow utf-8 input
\usepackage[T1]{fontenc}    % use 8-bit T1 fonts
\usepackage{hyperref}       % hyperlinks
\usepackage{url}            % simple URL typesetting
\usepackage{booktabs}       % professional-quality tables
\usepackage{amsfonts}       % blackboard math symbols
\usepackage{nicefrac}       % compact symbols for 1/2, etc.
\usepackage{microtype}      % microtypography
\usepackage{lipsum}		% Can be removed after putting your text content
\usepackage{graphicx}
\usepackage{url}

\title{IoT Blockchain Solution for Air Quality Monitoring in SmartCities}

\author{ \href{https://orcid.org/0000-0002-2543-2710.}{\includegraphics[scale=0.06]{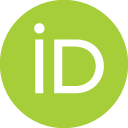}\hspace{1mm}Shajulin~Benedict}\thanks{(\url{www.iiitkottayam.ac.in/shajulin.php}---\emph{© IEEE ANTS 2019 Personal use of this material is permitted. Permission from IEEE must be obtained for all other uses, in any current or future media, including reprinting/republishing this material for advertising or promotional purposes, creating new collective works, for resale or redistribution to servers or lists, or reuse of any copyrighted component of this work in other works.}}) \\
	Faculty, Department of Computer Science\\
	Indian Institute of Information Technology Kottayam \\
	Kerala, India -- 686635. \\
	\texttt{shajulin@iiitkottayam.ac.in} \\
	\And
	\href{}{\includegraphics[scale=0.06]{orcid.png}\hspace{1mm}Rumaize P.} \\
	Student, Department of Computer Science\\
	Indian Institute of Information Technology Kottayam \\
	Kerala, India -- 686635. \\
	\texttt{rumaize@iiitkottayam.ac.in} \\
	\And
	\href{}{\includegraphics[scale=0.06]{orcid.png}\hspace{1mm}Jaspreet Kaur} \\
	Internee'19, Department of Computer Science\\
	Indian Institute of Information Technology Kottayam \\
	Kerala, India -- 686635. \\
}

\begin{document}
\maketitle

\begin{abstract}
	IoT cloud enabled societal applications have dramatically increased in the recent past due to the thrust for innovations, notably through startup initiatives, in various sectors such as agriculture, healthcare, industry, and so forth. The existing IoT cloud solutions have led practitioners or researchers to a haphazard clutter of serious security hazards and performance inefficiencies. This paper proposes a blockchain enabled IoT cloud implementation to tackle the existing issues in smart cities. It particularly highlights the implementation of chaincodes for air quality monitoring systems in SmartCities; the proposed architecture named as IoT enabled Blockchain for Air Quality Monitoring System (IB-AQMS) is illustrated using experiments. Experimental results were carried out and the findings were disclosed in the paper.  
\end{abstract}

% keywords can be removed

\section{Introduction}
IoT Cloud based technology \cite{Gubbi:2013:Journal} \cite{Lake:2014:Journal} is almost prevalent in most diverse research sectors, including agriculture, healthcare, societal environment, smartcities, smart manufacturing \cite{Lin:2016:Conf} \cite{Quiros:2017:Conf} \cite{Wang:2016:Journal} and so forth. Tens of thousands of researchers or industrialists, including startups, have invested research notions to inculcate the newer technologies on the existing solutions for reaping in profits or societal benefits. 

Although several IoT cloud solutions do exist in the market (including smart cities), they are prone to serious concerns relating to security and performance inefficiencies. Notably, IoT enabled air quality measurement systems for smart cities might pinpoint the industrial emissions to the concerned governmental agencies such as Europeon Environment Agency of Europe, Environmental Protection Agency (EPA) of the USA, or Central Pollution Control Board (CPCB) of India. The record might be tampered with corrupt policies or the concerned industrialists. Obviously, there is a dire need of a tamperless secured record keeping solution for smart city officials. In fact, blockchain technology \cite{Zheng:2011:Conf} has been developed in the recent past for various applications \cite{Hori:2018:Conf} \cite{Mauro:2018:Conf}. 

This paper proposes an IoT Blockchain enabled Air Quality Monitoring System (IB-AQMS) for smart cities. The proposed approach records the air pollution details of industries from a smart city in the form of untampered blocks. The sensor data are collected through IoT devices, and they are validated based on the underlying peers of blockchains. The implementation was validated at the IoT cloud research laboratory of IIIT Kottayam. And, the evaluation results were discussed. 

The rest of the paper is organized as follows: Section~\ref{sec:RelatedWork}
presents a wide survey on the topic of IoT blockchains for smart cities. Section~\ref{sec:architecture}
explains the proposed IoT Blockchain enabled Air Quality Monitoring System architecture. 
Section~\ref{sec:chaincodes} describes the inner details of transactions and chaincodes of blockchains. Section~\ref{sec:Experiment} reveals the experimental results and advantages of blockchain based implementation for smart cities. And,  
finally, Section~\ref{sec:Conclusion} presents a few conclusions. 

\section{Related Work} \label{sec:RelatedWork}
Smart city research and solutions have improved in the recent past owing to several innovative solutions or product developments. Solutions such as i) providing sufficient bandwidth via. 5G, ii) intelligent services in various sectors, including transportation and smart waste management \cite{Haribabu:2017:Conf}, \cite{Malapur:2017:Conf}, \cite{Bharadwaj:2017:Conf}, \cite{AlMasri:2018:Conf}, iii) controlling air/water pollution levels, iv) notification or alert services, and so forth have evolved in various dimensions in the recent past. 

\begin{figure*}[!h]
\centering
\includegraphics[width=1.0\columnwidth]{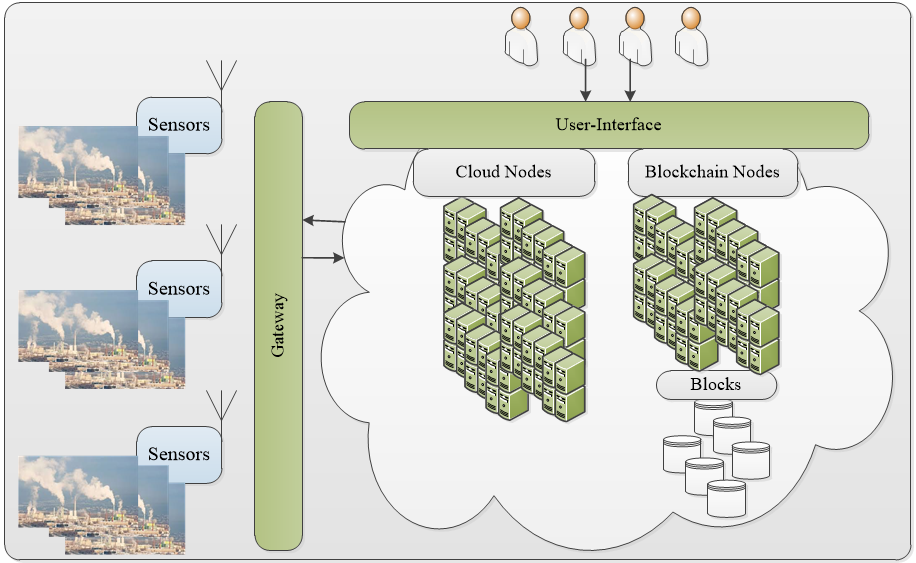} 
\caption{Architecture of IB-AQMS}
\label{fig:architecture}
\end{figure*}

In addition, city rankings  (~\cite{Giffner:2007:web,Rolland:2017:web}) and the other notable schemes such as Swatchh Bharath of India have promoted city authorities to adopt effective measures and policies to control wrong practices; a few indexing schemes do exists for ranking countries such as Air Quality Indexing \cite{AQI:2019:web}.    

Controlling air quality pollution level is carried out at outdoor, indoor, or at the industrial sectors. For instance, Kadri et al \cite{Kadri:2013:conf} proposed a machine-2-machine approach of monitoring air quality parameters in a distributed fashion. 
Also, Yash et al \cite{Yash:2016:conf} have proposed a cloud assisted air quality monitoring approach. Siemens has developed a software to reveal the air quality values of cities \cite{Siemens:2017:web}. 

A few researchers have studied the utilization of prediction approaches to study the variables such as NO2, SO2, RSPM, or so forth -- i.e., authors of \cite{Santhosh:2011:Conf} have studied the application of neural network based prediction algorithm for predicting 
air quality parameters; Pietro et al \cite{Pietro:2008:Journal} have analyzed the impact of the severity of CO and NO2 along roadsides using the neural network 
prediction algorithm; authors of \cite{Yunliang:2017:Journal} have explored the effect of utilizing EPLS method for clustering air quality data. In addition, 
authors of \cite{Cai:2016:Journal}, have created pure analytical models for analyzing the air quality of a specific region. 

In the previous work \cite{Benedict:2017:Conf}, a revenue oriented air quality prediction framework using Random Forest algorithm which has the capability of notifying control authorities and which creates revenue to the smartcities using cloud microservices were designed. 

In most of the available mechanisms, there is a possibility of industrialists or concerned air polluting entities to tamper the data. This tampering notion could heavily impact the control authorities of smartcities from leveraging strict policies or protecting the environment. The work, proposed in this paper, has endeavored to apply blockchain mechanism to register the measured air quality sensor data from various air quality measurement sites such that the generated blocks remain untampered from the defaulters. 

\section{IB-AQMS for SmartCities} \label{sec:architecture}
Air pollution is considered to be a serious issue, especially in developing countries. The air pollution leads to an array of health hazards, including pre-mature deaths, to the residents of SmartCities. Albeit of a dramatic decrease in the air pollutant emissions, there still exist issues due to a few wrong practices heralded by industrialists. 
This section details on the IoT Blockchain enabled Air Quality Monitoring System (IB-AQMS) for smart cities in order to avoid any tempering caused by malpractitioners at the SmartCity system.

The proposed architecture, IB-AQMS, consists of the following entities: 
\begin{enumerate}
\item Air quality sensors / Gateway
\item Cloud Nodes 
\item Blockchain Nodes
\item User-Interface
\end{enumerate}

The pictorial representation of the architecture is given in Figure~\ref{fig:architecture}. 

The functionalities of these entities are listed in the following subsections: 

\subsection{Air Quality Sensors / Gateway}
Sensors are typically battery operated devices which measure a few measurable properties such as temperature, NO2, SO2, CO2, dust, and so forth. These sensors are often incapable of handling larger requests or executing compute-intensive tasks. Hence, such devices are connected to an IoT gateway for performing the underlying tasks of applications. The important roles of the IoT gateway are listed below: 
\begin{itemize}
\item to establish connection to the external world via. proxy servers. 
\item to aggregate a large volume of sensors belonging to multiple vendors or multiple protocols such as IEEE802.15.4, Zigbee-P, RFID, Bluetooth4.0, and so forth. 
\item to provide unique addresses or geo-spacial addresses based on the context of applications, and 
\item to enrich the analytics of the underlying sensor data (if required) to a minimal level so that all sensor data are not transferred to the cloud environments. 
\end{itemize}

\begin{figure*}[!h]
\centering
\includegraphics[width=1.0\columnwidth]{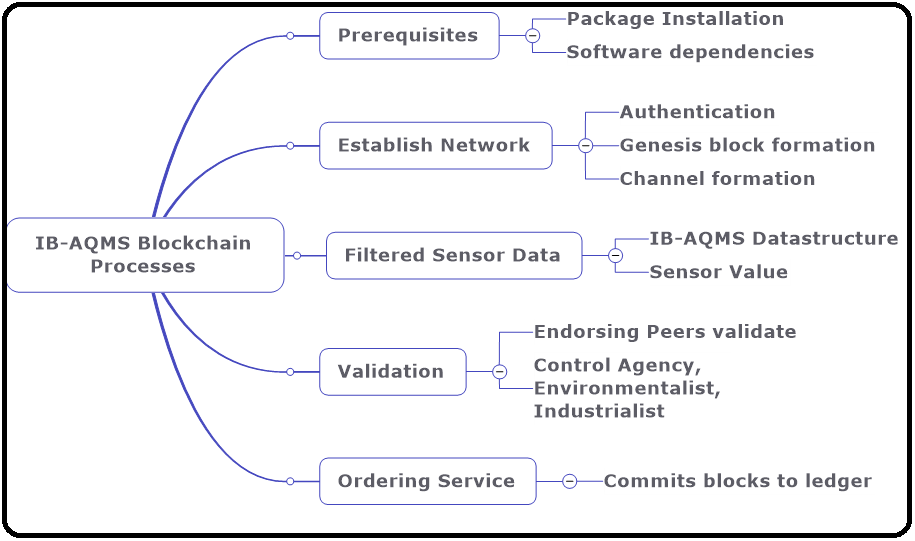} 
\caption{Blockchain Process Flow of IB-AQMS}
\label{fig:processes}
\end{figure*}

\subsection{Cloud Nodes} 
The sensor data, either filtered or the otherwise, reaches cloud environments. The cloud environments are typically provisioned by a few public cloud providers such as Amazon AWS, Google Compute Engine, or IBM Cloud, or by a few in-house opensource cloud setup based on OpenStack, OpenNebula or so forth. These cloud servers are computationally powerful to handle several functions: 

\begin{itemize}
\item data analytics -- an exploratory analysis of upcoming air quality sensor data is necessary to characterize the faulty industries which emit air pollutants to the environment. In this context, it is possible to adopt several data modeling or prediction algorithms or machine learning algorithms using cloud services; 
\item executing algorithms -- launching VM instances from the cloud templates or starting a new instance of a VM after contending for resources from cloud providers are primordial steps to execute algorithms or IoT applications. The cloud nodes would have to deal with all these underpinning tasks for executing IoT cloud algorithms, including blockchain services;
\item monitoring the events -- sometimes, there is a possibility that the resources are not powered on due to several reasons -- for instance, the accountability of the cloud user is not validated by the IAM cloud service. In such cases, the cloud nodes remain unavailable to execute IoT cloud tasks. Monitoring the performance efficiency of IoT application, therefore, is carried out at cloud nodes;
\item security -- additionally, cloud specific security measures such as cross site scripting, device hacking, VM rootkits, and so forth are handled at cloud nodes; and so forth.
\end{itemize} 

\subsection{Blockchain Nodes}
The blockchain nodes of IB-AQMS are, typically, a few VM instances or dedicated servers of cloud providers. Blockchain nodes are connected in a Peer-2-Peer distributed network fashion such that the nodes belong to various organizations. Each organization shall include multiple peer nodes where the copy of blocks are located. Each block contains the hashed values of the previous blocks and the transaction data along with the timestamp.  The most predominant services that we have opted in this work are the application of blockchain for IoT cloud applications. 

The pristine roles of these services are detailed in the following section \ref{sec:chaincodes}

\subsection{User-Interface}
The next entity of IB-AQMS architecture is the User-Interface. Clients or Users could login to the IB-AQMS system to check the status of blockchain data. Notably, if the data is hampered by any defaulters, the whole blockchain ledger fails in the operations. 

\section{Blockchain Nodes -- Working Principle} \label{sec:chaincodes} 
This section explains how blockchain nodes assist the process of adding sensor data to the ledger (see Figure~\ref{fig:processes}).

The blockchain nodes of IB-AQMS are categorized as follows: 
\begin{enumerate}
 \item \texttt{Actors} -- these nodes are responsible to initiate the transaction to the ledger. 
 \item \texttt{Endorsing Peers} -- such nodes that are utilized to verify the validity of transactions. 
 \item \texttt{Orderers} -- these nodes collect the validated transactions and issue them to the committing peers. 
 \item \texttt{Committing Peers} -- the nodes which register the transaction as blocks to the chain or ledger (which remains untampered) are named as \texttt{Committing Peers}. 
\end{enumerate}
All blockchain nodes follow certain rules based on \texttt{Chaincodes} -- also named as smart contract -- during the process of registering transactions to the ledger. The \texttt{Chaincodes} are the pieces of codes that are required to alter the state of the ledger. These \texttt{Chaincodes} are executed on independent nodes to the blockchain nodes.

\begin{figure*}[!h]
\centering
\includegraphics[width=.80 \columnwidth]{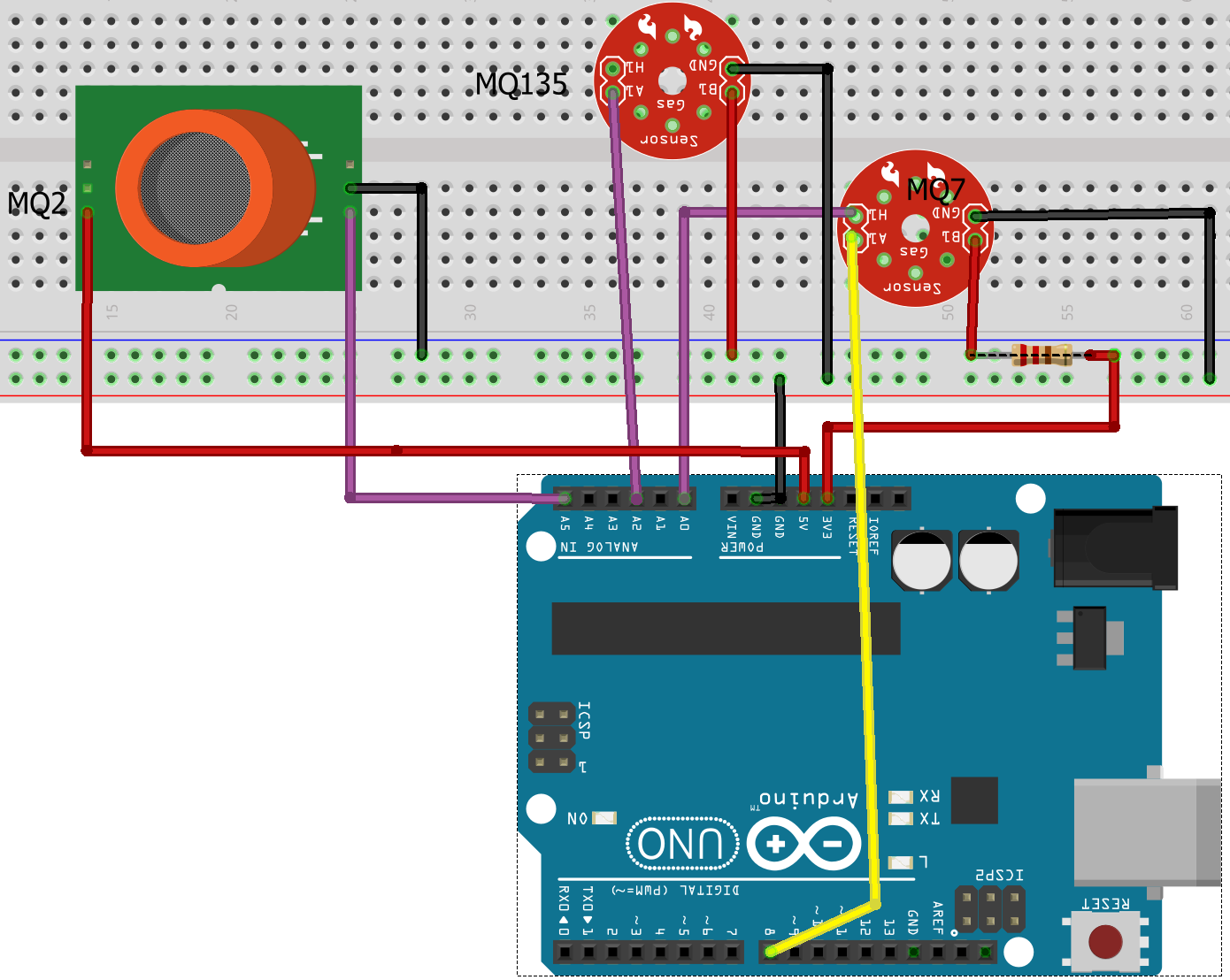} 
\caption{Circuit Diagram for sensing Air Quality Parameters such as CO, CO2, and Smoke particles}
\label{fig:circuit}
\end{figure*}

In IB-AQMS architecture, the blockchain services (see Figure~\ref{fig:processes}) follow a sequence of procedures in order to register the block into the ledger. They are listed as follows: 
\begin{enumerate} 
\item \textit{Establishing network} -- at first, all required pre-requisites are satisfied such as downloading the required software, packages, and so forth. And, the required authentication certificates and the genesis block (the first block) of a blockchain are created based on the specifications provided at the configuration file.  
\item \textit{Submitting Sensor Data} -- the sensors -- submit the sensory data to the blockchain service. The sensory data follow in a specific structure consisting of i) timestamp, ii) type of location, iii) SO2, iv) NO2, v) RSPM, vi) CO, vii) Associated Industry Names, viii) Monitoring location, ix) Penalty value, and x) Reporting Agency. These structured sensory data are preferably having the higher air pollutant emissions from a particular IoT monitoring site; 
\item \textit{Validation} -- upon the receipt of the sensory data, the Endorsing Peers, which are represented as \texttt{Control Agency}, \texttt{Environmentalist}, and \texttt{Industrialists}, look into the validation of data and the authenticity of the submitting client based on the chaincodes. 
\item \textit{Ordering Service} -- the transaction is passed on to the ordering service of the Orderer node once the previous step was completed. 
\item \textit{Committing Blocks} -- Accordingly, the Orderer lets the Committing Peers to register the information or block into the blockchain by adding the previous hash values. Although the blocks are maintained by Peers, they are not let to be tampered by any of the nodes due to the specialty of the blockchain network. 
\end{enumerate}

\section{Experimental Results} \label{sec:Experiment}
This section illustrates the experimental setup, findings of the experiments, and the corresponding discussions. All experiments were carried out at the IoT Cloud Research Laboratory of our premise. 

\subsection{Experimental Setup}
In fact, the real-time air-quality sensor values could be collected using air quality sensors through the WIFI enabled ArduinoUNO board. The sketches of ArduinoUNO programming could include a specific URI to upload the real time value of air quality sensor value to the cloud setup as carried out in the other works \cite{Benedict:2017:Conf}. 

The arduino circuit utilized for collecting the air quality values from sensors such as MQ2, MQ135, MQ7 for measuring CO, smoke, and CO2 values is given in Figure~\ref{fig:circuit}. 

As seen in Figure~\ref{fig:circuit}, the sensors were connected to the analog pins of the arduino boards A0, A2, and A5 using specific pin modes of the board -- 
\begin{verbatim} 
    pinMode(mq7Dpin,INPUT);
    pinMode(smokeA0,INPUT);
    pinMode(mq135val,INPUT); 
 \end{verbatim}
 
 The sensors are powered by the supply units of the board. The sensor values obtained during the experiments are given in Table~\ref{tab:sensorValues}.

\begin{table}[]
\caption{Air Quality Measurement Values using Sensors}
\centering
\begin{tabular}{|l|l|l|}
\hline
MQ-7 (CO) & MQ2 (Smoke) & MQ135 (CO2) \\ \hline
379       & 375         & 381         \\ \hline
388       & 380         & 389         \\ \hline
393       & 384         & 394         \\ \hline
395       & 385         & 394         \\ \hline
394       & 384         & 394         \\ \hline
393       & 382         & 392         \\ \hline
388       & 378         & 383         \\ \hline
380       & 371         & 380         \\ \hline
377       & 364         & 376         \\ \hline
373       & 357         & 373         \\ \hline
370       & 350         & 369         \\ \hline
365       & 342         & 364         \\ \hline
360       & 334         & 359         \\ \hline
355       & 326         & 355         \\ \hline
\end{tabular} \label{tab:sensorValues}
\end{table}

The focus of this work is to reveal the utilization of blockchain network for AQMS and manifest the possibility of tampering blocks.  

In this lieu, the IB-AQMS automatically collected the severity of air pollutant emissions such as SO2, NO2, RSPM/PM10, and PM2.5 from various monitoring sites based on the datasets \cite{Airquality:2019:Web}. IB-AQMS utilized an extended version of docker-based hyperledger fabric images (version 1.4.1) supported by golang v.1.11.5 and docker v18.09.6 in order to provide blockchain services. The name of the nodes utilized in the blockchain services and the associated peers of IB-AQMS, the entire organizational settings, are tabulated in Table \ref{tab:settings}.  

\begin{table*}[]
\centering
\caption{Experimental Settings of IB-AQMS Blockchain Services}
\begin{tabular}{|l|l|l|l|l|}
\hline
\multicolumn{5}{|l|}{\textbf{BlockChain Services of IB-AQMS}} \\ \hline
Org & Blockchain Nodes & Service URI & Network & Volumes \\ \hline
1 & Peer 0 & http://peer0.iiitkottayam.com:7051 & fibchannel & peer0.iiitkottayam.com \\ \hline
1 & Peer 1 & http://peer1.iiitkottayam.com:8051 & fibchannel & peer1.iiitkottayam.com \\ \hline
2 & Peer 0 & http://peer0.aic.com:9051 & fibchannel & peer0.aic.com \\ \hline
2 & Peer 1 & http://peer1.aic.com:10051 & fibchannel & peer1.aic.com \\ \hline
O & Orderer & http://orderer.iiitkottayam.com:7050 & fibchannel & orderer.iiitkottayam.com \\ \hline
\end{tabular} \label{tab:settings}
\end{table*}
Peer 0 of each organizations was responsible for endorsing transactions; and, iiitkottayam.com was utilized as \texttt{Orderer} throughout the experiments.

\begin{figure*}[]
\centering
\includegraphics[width=1\columnwidth]{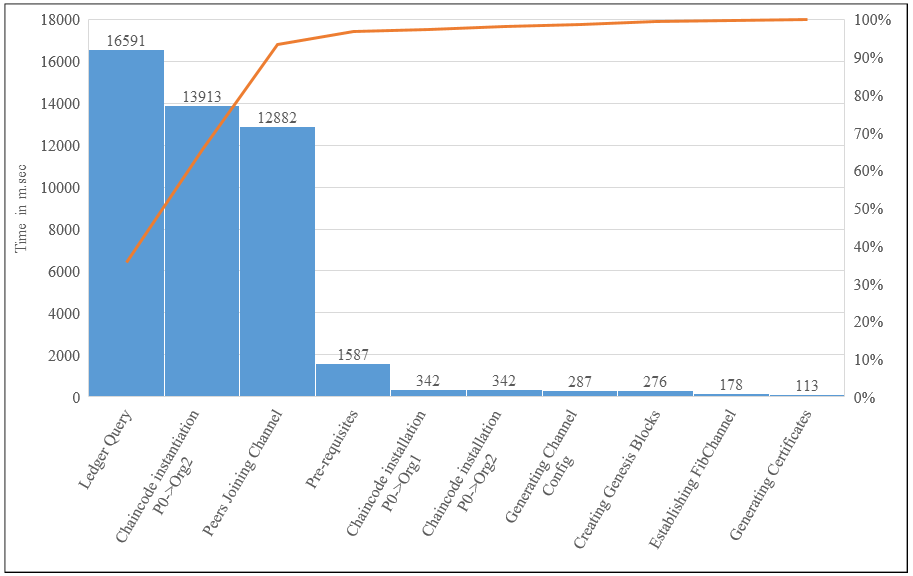} 
\caption{Execution Time of IB-AQMS Blockchain Processes}
\label{fig:exectime}
\end{figure*}

\subsection{Execution Time} 
The processes involved in registering the string of sensor data, a block representation of blockchain, into the ledger of IB-AQMS (as discussed in Section~\ref{sec:architecture}) were instrumented with specific monitoring hooks in order to  study the time spent by these processes on the organizational nodes. Figure \ref{fig:exectime} depicts on the impact of the processes, in terms of the execution time in milli-seconds, of blockchain services when executed on the the nodes.  

The following points are observed from the Figure~\ref{fig:exectime}: 
\begin{enumerate}
\item The query time, the chain code instantiation time, and the peer joining to the blockchain time, are comparatively very high when compared to the other tasks such as Chaincode installation, establishing the channel, generating certificates, and so forth. 
\item The pre-requisites are the time required to setup the blockchain nodes based on the docker images. The time will be higher if the required images are not available in the executing nodes. 
\item The chaincode instantiation time is higher than the chaincode installation time owing to waiting for connections from neighboring peer nodes. 
\end{enumerate}

\section{Conclusion} \label{sec:Conclusion}
Air quality monitoring using IoT solutions have been an interesting topic of research for industrialists, environmentalists, and researchers of SmartCities in the recent past. Albeit of several solutions in the recent past, there have been a very few solutions that highlighted the security aspects of sensor data. Notably, air quality monitoring data might be destroyed or altered by any unethical industrialists when connected with a few corrupt controlling authorities. This paper proposed IB-AQMS architecture which combined blockchain technology to IoT enabled air quality monitoring systems. The approach was experimented using air quality monitoring sensors and the available datasets.  

% \section{Acknowledgements}
% This work is partially funded by the BEL consultancy project of IIIT-Kottayam. In addition, the authors thank IIIT-Kottayam officials, including Director, and AIM team for motivating towards the entrepreneuial journey in India. 

\end{document}